# Terahertz Spin–Transfer Torque Oscillator Based on a Synthetic Antiferromagnet

Hai Zhong[1], Shizhu Qiao[2, †], Shishen Yan[1], Hong Zhang[3], Yufeng Qin[3], Lanju Liang[2], Dequan Wei[2], Yinrui Zhao[1], Shishou Kang[1, ‡]

[1] School of Physics, State Key Laboratory of Crystal Materials, Shandong University, Jinan 250100, P. R. China

[2] School of Opt-Electronic Engineering, Zaozhuang University, Zaozhuang 277160, P. R. China

[3] Department of Applied Physics, School of Information Science and Engineering, Shandong Agricultural University, Taian 271018, P. R. China

[†]Corresponding author: ryjqyears@gmail.com

[‡]Corresponding author: skang@sdu.edu.cn

## Abstract

Bloch-Bloembergen-Slonczewski equation is adopted to simulate magnetization dynamics in spin-valve based spin-transfer torque oscillator with synthetic antiferromagnet acting as a free magnetic layer. High frequency up to the terahertz scale is predicted in synthetic antiferromagnet spin-transfer torque oscillator with no external magnetic field if the following requirements are fulfilled: antiferromagnetic coupling between synthetic antiferromagnetic layers is sufficiently strong, and the thickness of top (bottom) layer of synthetic antiferromagnet is sufficiently thick (thin) to achieve a wide current density window for the high oscillation frequency. Additionally, the transverse relaxation time of the free magnetic layer should be sufficiently larger compared with the longitudinal relaxation time. Otherwise, stable oscillation cannot be sustained or scenarios similar to regular spin valve-based spin-transfer torque oscillator with relatively low frequency will occur. Our calculations pave a new way for exploring THz spintronics devices.

## 1. Introduction

Terahertz (THz) technology, with its frequency ranging approximately within $10^{11}$–$10^{13}$ Hz, has numerous applications, such as high-resolution imaging, nuclear fusion plasma diagnosis, skin cancer screening, large-scale integrated circuit testing, weapons detecting, wireless communication, etc. [1-3]. Typical THz sources based on quantum cascade laser, superconductor Josephson junctions, electron tubes, accelerators, and other solid-state electronic devices have limitations of complex setups or low temperature, etc. [4,5], hindering size shrinking down and consequently, many potential applications.

However, fabricated by microscale or nanoscale technologies, spin transfer torque (STT) oscillator could shrink down to micrometer or sub-micrometer size [6-8]. Moreover, synthetic antiferromagnets (SAFs), consisting of two or more antiferromagnetic coupled ferromagnets separated by metallic spacers or tunnel barriers [9], are attracting increasing attention in STT spintronics [10-17], because, comparing with crystal antiferromagnets, they have considerably weak exchange interactions and notably larger spatial scales, leading to an easy manipulation of antiferromagnetic order [18]. Magnetization switching by spin-orbit torque is reported in MgO/CoFeB/Ta/CoFeB/MgO structure with two CoFeB layers in antiferromagnetic coupling through the Ruderman–Kittel–Kasuya–Yosida (RKKY) interaction, which may help reducing power consumption in magnetic memories with low stray field [16]. Compared with single-layer free magnetic layer, STT oscillation in SAF has higher output power and narrower linewidth [15].

Although acting as effective magnetic field, RKKY interaction may boost magnetization oscillation frequency greatly into THz range [5,19,20], no results on THz oscillation frequency has been reported to date in STT oscillator based on SAF, neither experimentally nor theoretically. In this paper, the Bloch–Bloembergen–Slonczewski (BBS) equation is adopted in spin-valve-based STT oscillator with SAF acting as the free magnetic layer, and THz oscillation frequency is achieved with reasonable parameters under no applied magnetic field. Simulation shows that, in order to achieve SAF STT oscillator with top frequency up to THz region, antiferromagnetic coupling between SAF layers should be sufficiently strong, and moreover, large (small) thickness of top (bottom) layer of SAF favors broad current density window of stable oscillation.

## 2. Theoretical Model

Typically, coupled macro-spin Landau–Lifshitz–Gilbert (LLG) equations, where the magnitude of magnetization keeps unchanged, are adopted to study magnetization dynamics of two magnetic layer of SAF [21,22]. However, short–wavelength magnon excitation, diminishing magnetization [23,24], plays an important role in magnetic relaxation, especially in thin film structures [25-28]. Moreover, all types of magnons, with wavelengths short and long, are the concern of magnon spintronics [29-31]. Consequently, macro-spin theoretical method based on LLG equation cannot deal with short–wavelength magnon excitation. On the contrary, the Bloch–Bloembergen equation [32-34],

$$\frac{d\boldsymbol{M}}{dt} = \gamma\mu_0 \boldsymbol{M} \times \boldsymbol{H}_{\text{eff}} - \frac{M_x}{T_2}\boldsymbol{e}_x - \frac{M_y}{T_2}\boldsymbol{e}_y - \frac{M_z - M_S}{T_1}\boldsymbol{e}_z, \qquad (1)$$

with two magnetization relaxation parameters, longitudinal relaxation time $T_1$, and transverse relaxation time $T_2$, rather than one Gilbert damping constant, is more flexible, and short–wavelength magnon excitation is included in $T_2$. For more information about the contribution of short-wavelength magnon excitation to magnetic relaxation, please refer to Ref. [35]. Bloch–Bloembergen–Slonczewski (BBS) equation has been established to study magnetic precession in regular STT oscillator with a single magnetic layer acting as the free magnetic layer [35], which demonstrated that $T_2 > T_1$ is crucial for stable STT oscillation.

In this study, we apply two coupled BBS equations to a spin valve structure with SAF acting as the free magnetic layer. The BBS equation reads as

$$\frac{dM}{dt} = \gamma\mu_0 M \times H_{eff} - \frac{M_x}{T_2}e_x - \frac{M_y}{T_2}e_y - \frac{M_z - M_s}{T_1}e_z + \tau_{STT}, \qquad (2)$$

where $\tau_{STT}$ is the Slonczewski torque [36,37]. The spin valve with F0/N0/F1/N1/F2 structure is shown in Fig. 1. F0 is the fixed magnetic layer, with its magnetization pinned along the –z direction. Ferromagnetic layers F0 and F1 are separated by normal metal layer N0. Ferromagnetic layers F1 and F2, with thickness of $d_1$ and $d_2$, are antiferromagnetically coupled by RKKY exchange interaction through normal metal layer N1, and the F1/N1/F2 SAF acts as the free magnetic layer of the STT oscillator. To begin the calculation, we assume F1 and F2 both have uniaxial magnetic anisotropy along z axis. The current is along the –x direction, perpendicular to film plane.

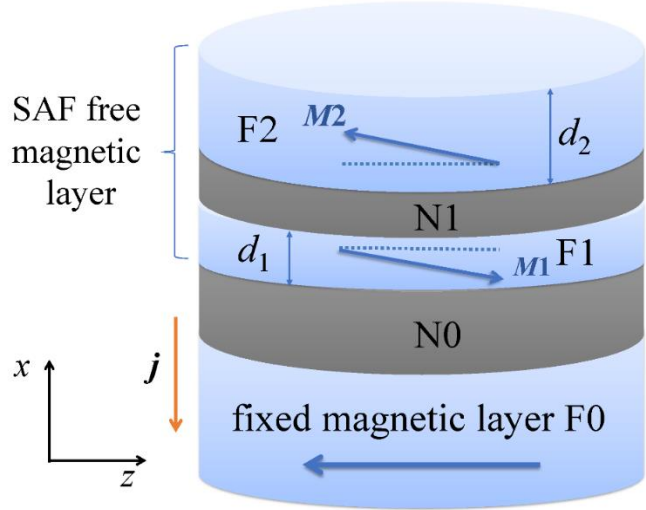

Fig. 1. Structure of spin valve with synthetic antiferromagnet (F1/N1/F2) acting as the free magnetic layer used in this study.

## 3. Simulation and Discussion

The following simulations and discussions are based on Fig. 1 and Eq. (2), and the parameters used in the simulations are listed in Table 1 unless otherwise specified. Parameters related to SAF are based on Co/Ru/Co [14].

Table 1. Parameter values used in simulations, where $M_s$, $H_K$, $d_1$ and $d_2$ are saturated magnetization, magnetic anisotropy field, and thickness of F1 and F2, respectively. $J_{ex}$ is the antiferromagnetic exchange parameter between F1 and F2. $P$ is the spin polarization of current.

| Parameter | Value | Parameter | Value |
| --- | --- | --- | --- |
| $M_s$ | $1.42 \times 10^6$ A/m | $J_{ex}$ | $-5 \times 10^{-3}$ J/m$^2$ |
| $H_K$ | $9.09 \times 10^4$ A/m | $P$ | 0.3 |
| $d_1$ | 2 nm | $T_2$ | 0.75 ns |
| $d_2$ | 8 nm | $T_1$ | 0.5 ns |

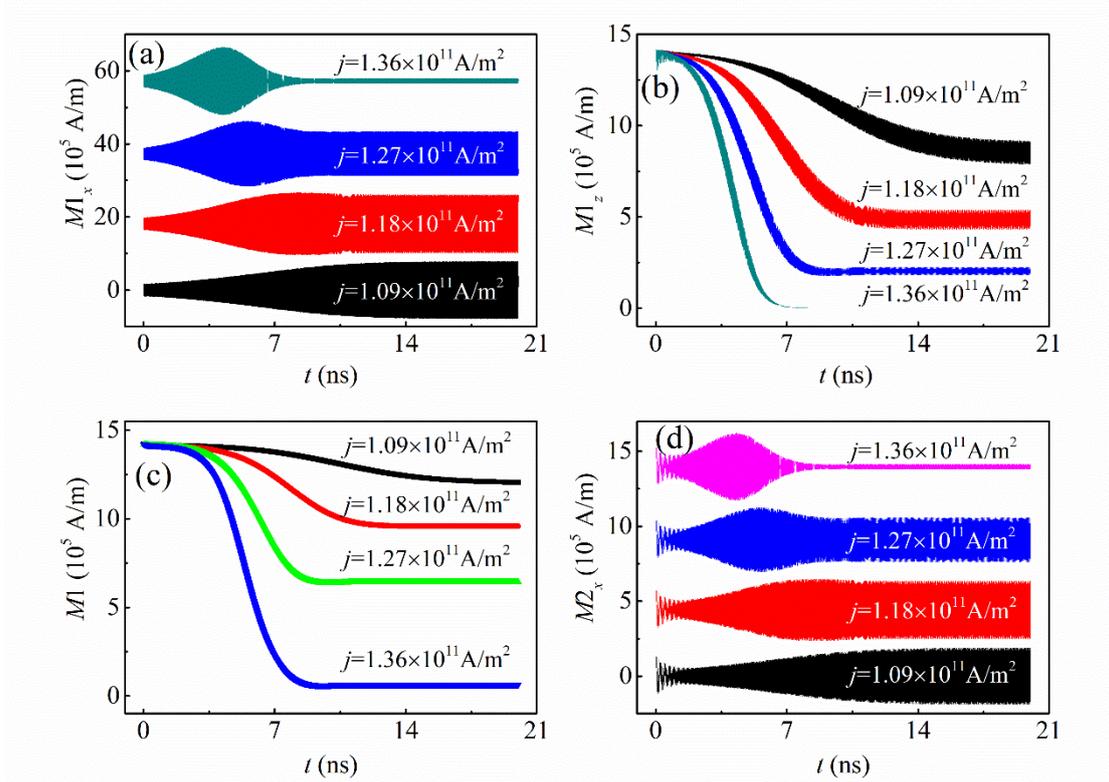

Fig. 2 Stable magnetization oscillation represented by evolutions of (a) $x$ component, (b) $z$ component of magnetization $M1$ of F1, (c) magnitude of $M1$, and (d) $x$ component of magnetization $M2$ of F2, where lines in (a) and (d) are stacked by vertical axis offset for clearness.

Stable magnetization oscillation of the free magnetic layer, which is the prerequisite of STT oscillator, could be achieved. As seen in Fig. 2(a) and (b), stable oscillations of *x* and *z* components of F1 magnetization is reached within less than 20 ns at various current densities from $1.09 \times 10^{11}$ A/m$^2$ to $1.36 \times 10^{11}$ A/m$^2$ (the same significant digits protocol for current density is kept in the following discussions). It seems quite odd that the amplitudes of these two oscillations decrease with the increasing of current density. This is caused by the fact that the density of short-wavelength magnons increases with the increase in current density, and short-wavelength magnons diminish the magnitude of magnetization [23,24], as seen in Fig. 2(c). Similar oscillation of *x* component of F2 magnetization is demonstrated in Fig. 2(d).

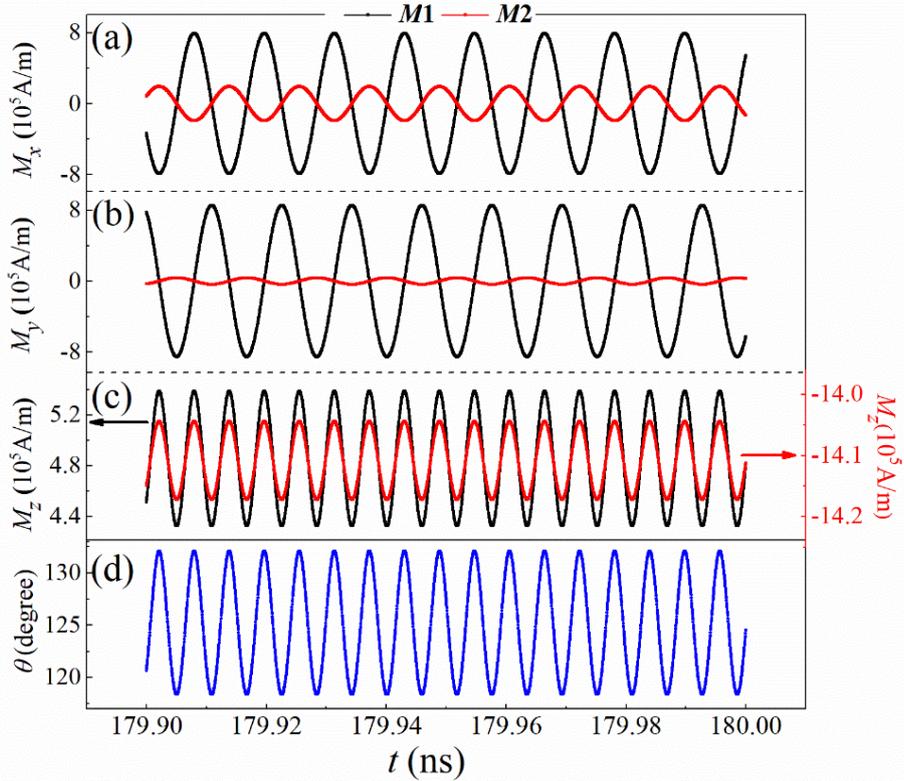

Fig. 3 Magnetization oscillation phases of F1 and F2 when $j = 1.18 \times 10^{11}$ A/m$^2$: (a) $M_x$, (b) $M_y$, (c) $M_z$, and (d) $\theta$ the angle between **M**1 and **M**2.

Phases in stable magnetization oscillation when $j = 1.18 \times 10^{11}$ A/m$^2$ is depicted in Fig. 3. The transverse part ($M_x$ and $M_y$) of $\boldsymbol{M}1$ and $\boldsymbol{M}2$ are opposite in phase as seen in Fig .3(a) and 3(b), and on the contrary, the longitudinal part, oscillating with twice the frequency of the transverse part, are in the same phase as seen in Fig. 3(c). $\theta$, the angle between $\boldsymbol{M}1$ and $\boldsymbol{M}2$, is shown in Fig. 3(d), which oscillates with the same frequency of the longitudinal part, ranging from 118° to 132°. The oscillation frequency mentioned in the following means the frequency of transverse part of magnetization, half the frequency of $M_z$ and $\theta$.

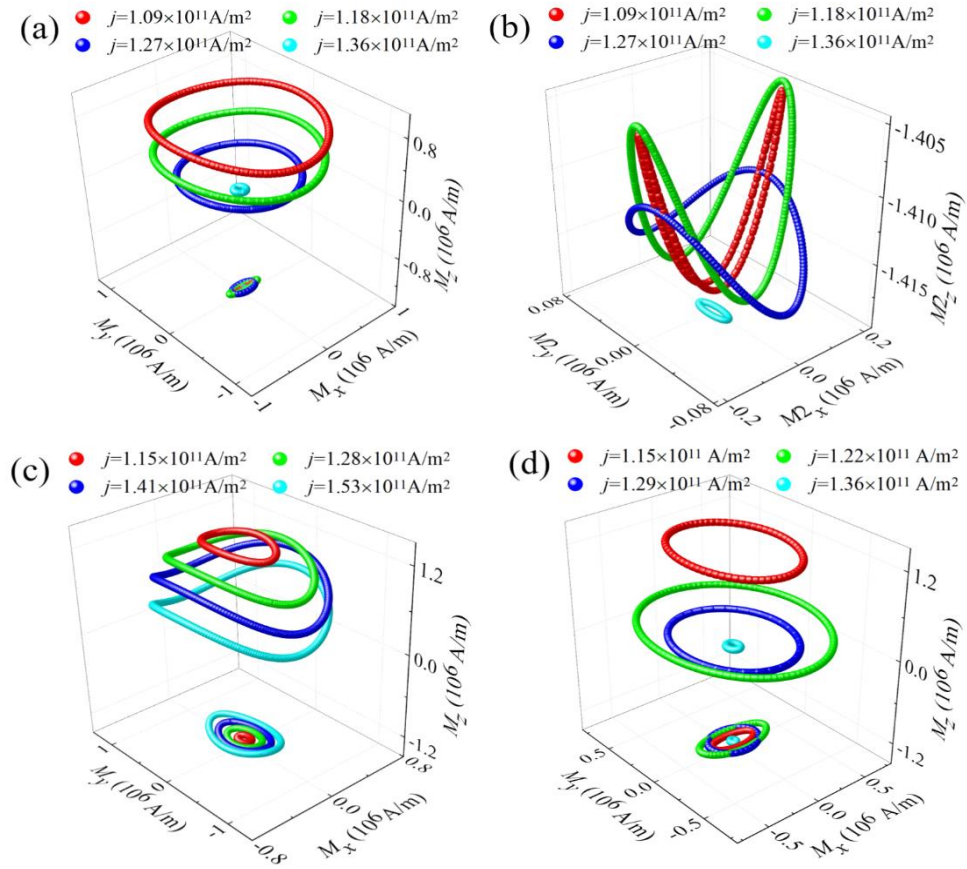

Fig. 4 (a) Magnetization trajectories of F1 (up) and F2 (down) at different current densities, (b) details of magnetization trajectories of F2 in (a). Magnetization trajectories of F1 (up) and F2 (down) when (c) the antiferromagnetic exchange parameter in Table 1 is replaced by $J_{\text{ex}} = -0.7$ J/m$^2$ and (d) the thickness of F2 layer in Table 1 is replaced by $d_2 = 4$ nm.

Fig. 4(a) demonstrates the magnetization trajectories of F1 and F2, which characterize two features. First, magnetization trajectories are not confined to a spherical boundary surface, different from the conclusion of macro-spin LLG model [36]. This is caused by the short-wavelength magnon excitation, which could also be observed in Fig. 2(c). When $j=1.36 \times 10^{11}$ A/m$^2$, magnetization trajectory is confined in a small space, and this short-wavelength magnon excitation-induced oscillation weakness should be well realized in STT oscillator application, because the output power of the oscillator is related to oscillation amplitude [38].

Secondly, oscillation amplitude of F2, whose oscillation details are shown in Fig. 4(b), is significantly smaller than that of F1. In addition, this phenomenon is revealed in Fig. 2(a), 2(d) and Fig. 3(a), 3(b), 3(c). The first reason behind this phenomenon is the exchange interaction between F1 and F2. This interaction tends to magnetize F2 antiparallel to F1, then the spin current passed through F1 contains a relatively small component that is transverse to the magnetization of F2, leading to a small amplitude of F2 oscillation. If the exchange interaction is weakened, the amplitude of F2 should be larger, which is indeed the simulation result shown in Fig. 4(c), where the antiferromagnetic exchange parameter $J_{ex} = -0.7$ J/m$^2$. The second reason is related to the thickness of F1 and F2. The saturation magnetizations and the cross-sectional sizes of F1 and F2 are set as the same in simulations, where the thickness of F1, $d_1 = 2$ nm, and the thickness of F2, $d_2 = 8$ nm. Under such a circumstance, the oscillation amplitude of F2 is smaller than that of F1 even though that the same amount of transvers spin current is absorbed by each of them. This is demonstrated in Fig. 4(d), where $d_1 = 2$ nm and $d_2 = 4$ nm, and comparing with Fig. 4(a), the oscillation amplitude of F2 is considerably larger.

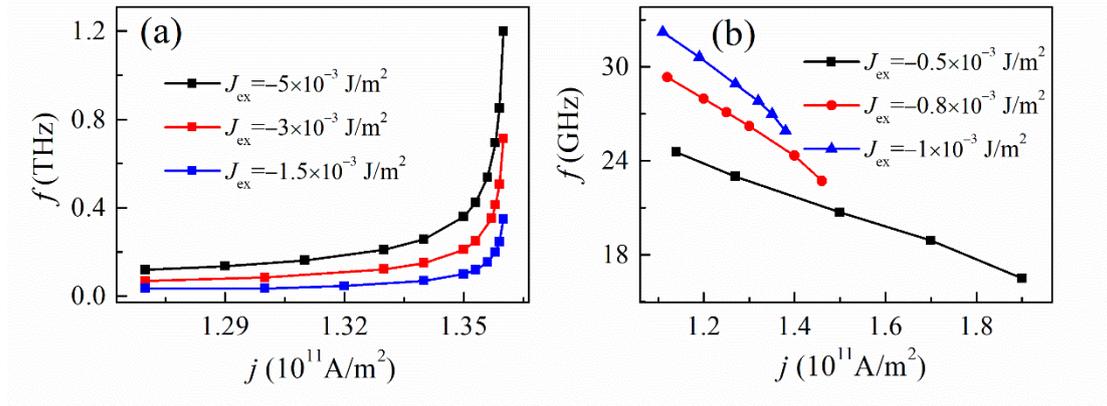

Fig. 5 Dependence of the $M_x$ frequency of stable magnetization oscillation on current density when antiferromagnetic exchange coupling between F1 and F2 is relatively (a) strong and (b) weak.

Since the magnetization oscillation amplitude of F1 is notably stronger than that of F2, in the following discussion, we will focus on the magnetization dynamics of F1. The oscillation frequency dependence on current density is shown in Fig. 5(a). Different from regular spin-valve based oscillator [35], the oscillation frequency of F1 in spin-valve-based SAF oscillator increases with the increase in current density in a distinct nonlinear manner, described as follow. In the relatively broad and weak current density region, the frequency increasing is mild, while in the other relatively narrow and strong current density region, the frequency increasing is acute, and the stronger the current density is, the more acute the frequency increase becomes. The frequency surpasses 1 THz (the frequency of $M_z$ and $M$ can surpass 2 THz) when $J_{ex} = -5 \times 10^3$ J/m$^2$ and $j = 1.36 \times 10^{11}$ A/m$^2$. This is because, acting as effective magnetic field, RKKY exchange interaction between F1 and F2 could boost the magnetization oscillation frequency to a large extent [19]. Moreover, the increase in the exchange interaction between F1 and F2, which could be realized by adjusting the thickness of normal metal layer N1 between F1 and F2 [9,39], leads to a higher oscillation frequency and wider frequency window.

Another interesting scenario occurs when the exchange interaction is too weak, as seen in Fig. 5(b). The oscillation frequency decreases with the increasing of current density, similar to the situation of regular spin-valve based oscillator [35], although the former has a slightly higher frequency owing to the additional exchange field in SAF. Similar change of blue shift to red shift of oscillation frequency dependence on current density due to applied field is reported in the studies based on LLG equation [22,40]. From Fig. 5, it can be observed that it is important to have strong RKKY exchange interaction between F1 and F2 to achieve a working THz SAF STT oscillator. It should be pointed out that RKKY exchange coupling is sensitive to the thickness of the normal metal layer in SAF, and whose fluctuation in space may compromise the performance of THz SAF STT oscillator [41], which may be one of the reasons why the THz SAF oscillator is difficult to achieve experimentally.

Although a high current density leads to a high frequency, even to THz zone, when $J_{ex}$ is sufficiently large, the oscillation amplitude is weakened sharply when the current density approaches the maximum value for the excitation of short-wavelength magnons, as seen in Fig. 2(a) and Fig. 4(a). This phenomenon leads to a sharp decrease of output power while approaching the maximum oscillation frequency and it seems inevitable: as seen in Fig. 4(c), we could increase the oscillation amplitude of the maximum current density ($j = 1.53 \times 10^{11}$ A/m$^2$) by weakening the RKKY exchange interaction ($J_{ex} = -0.7$ J/m$^2$). However, as seen in Fig. 5(b), weak exchange interaction leads to low oscillation frequency. This sharp decrease of output power may be another reason behind the difficulty in observing the THz oscillation experimentally, considering the small output power of the STT oscillator. Neglecting the excitation of short-wavelength magnons could be one of the reasons behind the absence of prediction for THz oscillation frequency by macro-spin LLG equation simulation, because wave energy is proportional to the square of frequency when the amplitude is fixed, which implies that the energy input to oscillator is 400 times larger for 1-THz oscillation compared with the 50-GHz oscillation in the macro-spin LLG simulation.

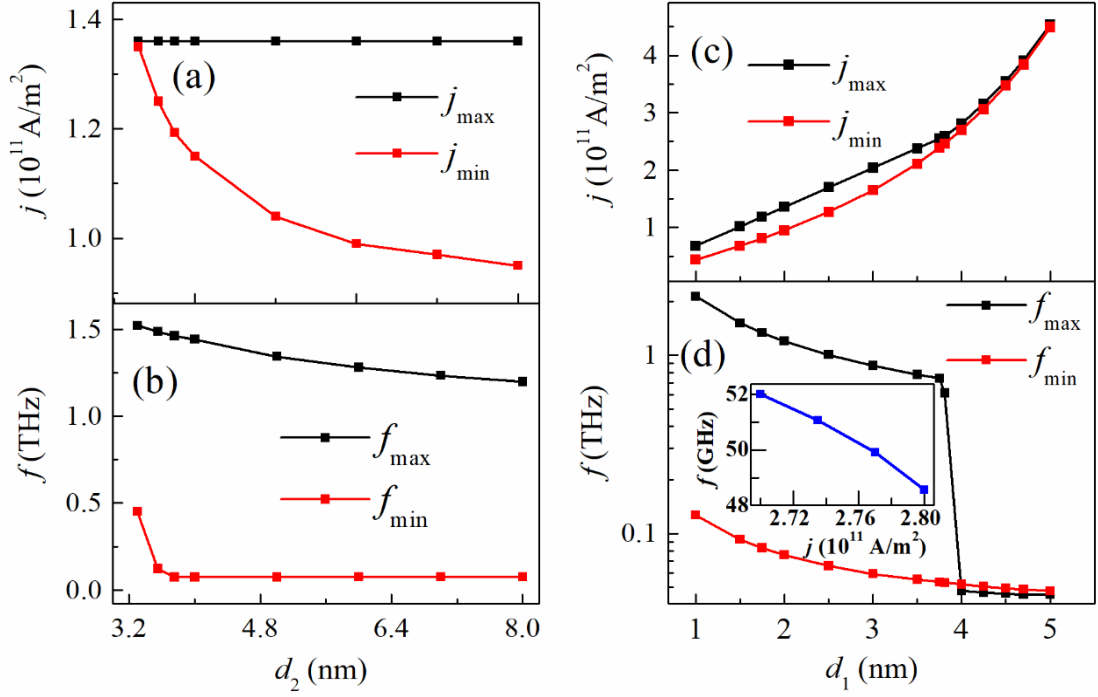

Fig. 6 Dependence of maximum and minimum current densities ($j_{max}$, $j_{min}$) for stable magnetization oscillation, and corresponding maximum and minimum oscillation frequencies ($f_{max}$, $f_{min}$) on the thickness of F2 (a), (b) and F1 (c), (d). The inset of (d) shows F1 oscillation frequency dependence on current density when $d_1 = 4$ nm and $d_2 = 8$ nm.

The effects of thickness of F2 and F1 on the magnetization oscillation are depicted in Fig. 6. The influence of $d_2$ on maximum and minimum current densities ($j_{max}$, $j_{min}$) for stable magnetization oscillation, and corresponding maximum and minimum oscillation frequencies ($f_{max}$, $f_{min}$) are shown in Fig. 6 (a) and (b). With the decrease in $d_2$, where $d_1 = 2$ nm, $j_{max}$ keeps unchanged within two significant digits after decimal point, while $j_{min}$ increases rapidly, which leads to the dramatic shrinking of current density window ($j_{max} - j_{min}$) and when $d_2 < 3.3$ nm, stable oscillation cannot be achieved. On the contrary, $f_{min}$ barely changes when $d_2 \geqslant 3.75$ nm, while $f_{max}$ increases evidently with the decreasing of $d_2$. $f_{min}$ increases dramatically when $d_2 < 3.75$ nm, leading to the shrinking of frequency window ($f_{max} - f_{min}$) under this circumstance.

The impact of variation in $d_1$ while $d_2 = 8$ nm is different to an extent. As seen in Fig. 6(c), both $j_{max}$ and $j_{min}$ increase with the increasing of $d_1$, and the current density window is maximized when $d_1 = 2.5$ nm. The current density window is approaching 0 when $d_1 > 4$ nm. As seen in Fig. 6(d) with the logarithmic vertical coordinate axis, $f_{min}$ does not vary considerably and $f_{max}$ decreases slower and slower with the increase in $d_1$ when $d_1 < 3.75$ nm, but it drops sharply to ~50 GHz when $d_1 > 3.75$ nm, and peculiarly, $f_{max} < f_{min}$ in this circumstance. This is further illustrated in the inset of Fig. 6(d), where the oscillation frequency of ~50 GHz decreases with the increase in the current density, similar to the weak exchange interaction scenario in Fig. 5(b). The exchange magnetic field in F1 is proportional to $J_{ex}/d_1$ [14], which may partially explain why the large $d_1$ scenario is similar to the weak exchange interaction scenario.

Consequently, Fig. 6 indicates that, in order to achieve better SAF oscillators, the thickness of F2 should be sufficiently large to broaden the current density window, and with the consideration of current density window and magnetic anisotropy, the thickness of F1 should be sufficiently small to increase $f_{max}$ and also to broaden the frequency window. It should be pointed out that some error in $f_{max}$ may occur due to our significant digits protocol and the sharp increase in frequency when the current density approaching maximum value, as seen in Fig. 5(a). We ran the simulations under a higher standard of significant digits protocol when $d_1 > 3.75$ nm, but no significant difference has been found. Even if high frequency was achievable when $d_1 > 3.75$ nm, the tuning of high frequency by current is very inconvenient, and may even be impossible, because a sharp increase in frequency occurs in a very narrow range of current density.

Similar to the scenario of the regular-spin-valve-based STT oscillator, the short-wavelength magnon excitation, contributing to transverse relaxation time $T_2$, has crucial influences on magnetization oscillation. Fig. 7(a) depicts the influence of $T_2$ on the maximum and minimum current densities required for stable magnetization oscillation. In this case, the current density window broadens dramatically, which favors the manipulation of SAF STT oscillator, with the increase in $T_2$, and it shrinks to zero if $T_2 < 0.52$ ns, which means that, under this circumstance, stable magnetization oscillation cannot be reached.

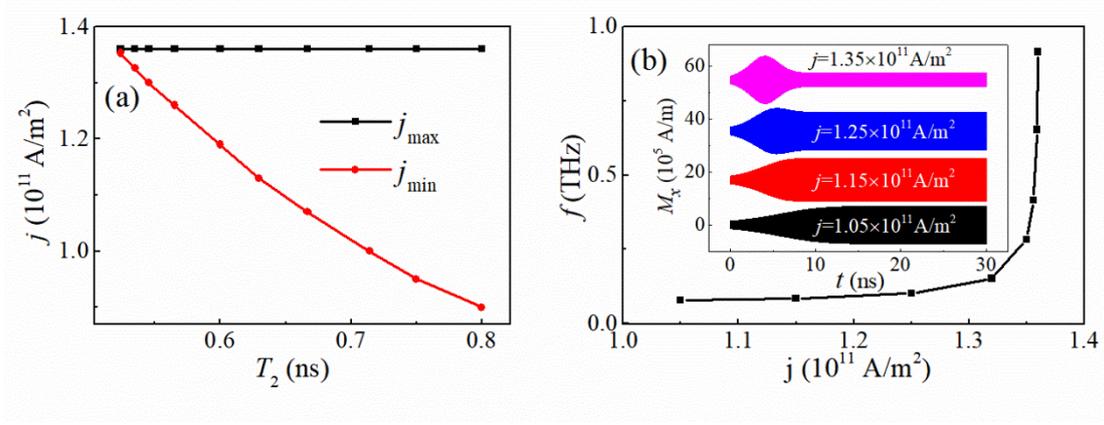

Fig. 7 (a) Influence of transverse relaxation time $T_2$ on maximum and minimum current densities ($j_{max}$ and $j_{min}$) for stable magnetization oscillation, (b) F1 stable oscillation frequency dependence on current density when F2 is pinned along the $-z$ direction; the inset in Fig. 7(b) shows the evolution of F1 magnetization of different current densities.

Finally, we should point out that no significant difference happens to F1 when F2 is pinned along the $-z$ direction. The frequency dependence on the current density in stable oscillation of F1 in this circumstance is shown in Fig. 7(b), which also clearly shows a nonlinear behavior. The evolution of magnetization with different current densities are depicted in the inset of Fig. 7(b), which shows a behavior similar to that demonstrated in Fig. 2(a).

## 4. Conclusion

In summary, considering the short-wavelength magnon excitation, we simulate magnetization oscillation in spin-valve based spin-transfer torque oscillator with a synthetic antiferromagnet acting as the free magnetic layer by coupled macro-spin Bloch–Bloembergen–Slonczewski equations. Magnetization oscillation frequency increases with the increasing of current density in a nonlinear manner to terahertz region when an antiferromagnetic coupling between the two synthetic antiferromagnet layers is relatively strong, which, however, decreases with the increase in the current density in a nearly linear manner when the coupling is too weak. The thickness of F2 (F1) should be sufficiently large (small) to achieve a better oscillator performance. Comparing with the longitudinal relaxation time, the transverse relaxation time of free magnetic layers should be sufficiently larger in order to achieve broad current density window of stable oscillation. The thickness fluctuation of normal metal layer in synthetic antiferromagnet, the sharp decrease in output power while approaching the terahertz frequency range, and the lack of consideration of short-wavelength magnon excitation may be among the reasons behind the absence of experimental observation and macro-spin LLG theoretical prediction of terahertz oscillation frequency. Spin-transfer torque oscillator based on synthetic antiferromagnet may shrink the terahertz oscillator and broaden its applications.

## ACKNOWLEDGEMENTS

This research is supported by the National Basic Research Program of China (Grant No. 2015CB921502), the key program of NSFC No. 11434006, and NSFC Nos. 11647034, 11674187, 11474184, and 61735010, the Natural Science Foundation of Shandong Province (Grant No. ZR2017MF005).

**References**

[1] T. Seifert *et al.*, Efficient metallic spintronic emitters of ultrabroadband terahertz radiation, *Nature Photonics* **10**, 483 (2016).

[2] M. Tonouchi, Cutting-edge terahertz technology, *Nature Photonics* **1**, 97 (2007).

[3] L. Consolino, S. Jung, A. Campa, M. De Regis, S. Pal, J. H. Kim, K. Fujita, A. Ito, M. Hitaka, S. Bartalini, P. De Natale, M. A. Belkin, and M. S. Vitiello, Spectral purity and tunability of terahertz quantum cascade laser sources based on intracavity difference-frequency generation, *Science Advances* **3**, e1603317 (2017).

[4] M. Hangyo, Development and future prospects of terahertz technology, *Jpn. J. Appl. Phys.* **54**, 120101 (2015).

[5] O. R. Sulymenko, O. V. Prokopenko, V. S. Tiberkevich, A. N. Slavin, B. A. Ivanov, and R. S. Khymyn, Terahertz-Frequency Spin Hall Auto-oscillator Based on a Canted Antiferromagnet, *Phys. Rev. Appl.* **8**, 064007 (2017).

[6] A. Sharma, A. A. Tulapurkar, and B. Muralidharan, Resonant Spin-Transfer-Torque Nano-Oscillators, *Phys. Rev. Appl.* **8**, 064014 (2017).

[7] A. A. Awad, P. Dürrenfeld, A. Houshang, M. Dvornik, E. Iacocca, R. K. Dumas, and J. Åkerman, Long-range mutual synchronization of spin Hall nano-oscillators, *Nat. Phys.* **13**, 292 (2016).

[8] Z. Duan, A. Smith, L. Yang, B. Youngblood, J. Lindner, V. E. Demidov, S. O. Demokritov, and I. N. Krivorotov, Nanowire spin torque oscillator driven by spin orbit torques, *Nat. Commun.* **5**, 5616 (2014).

[9] S. S. P. Parkin and D. Mauri, Spin engineering: Direct determination of the Ruderman-Kittel-Kasuya-Yosida far-field range function in ruthenium, *Phys. Rev. B* **44**, 7131 (1991).

[10] M. Romera, E. Monteblanco, F. Garcia-Sanchez, B. Delaët, L. D. Buda-Prejbeanu, and U. Ebels, Influence of interlayer coupling on the spin-torque-driven excitations in a spin-torque oscillator, *Phys. Rev. B* **95**, 094433 (2017).

[11] Y. C. Lau, D. Betto, K. Rode, J. M. Coey, and P. Stamenov, Spin-orbit torque switching without an external field using interlayer exchange coupling, *Nat. Nanotechnol.* **11**, 758 (2016).

[12] S. H. Yang, K. S. Ryu, and S. Parkin, Domain-wall velocities of up to 750 m s(-1) driven by exchange-coupling torque in synthetic antiferromagnets, *Nat. Nanotechnol.* **10**, 221 (2015).

[13] H. Saarikoski, H. Kohno, C. H. Marrows, and G. Tatara, Current-driven dynamics of coupled domain walls in a synthetic antiferromagnet, *Phys. Rev. B* **90**, 094411 (2014).

[14] C. Klein, C. Petitjean, and X. Waintal, Interplay between nonequilibrium and equilibrium spin torque using synthetic ferrimagnets, *Phys. Rev. Lett.* **108**, 086601 (2012).

[15] D. Houssameddine, J. F. Sierra, D. Gusakova, B. Delaet, U. Ebels, L. D. Buda-Prejbeanu, M. C. Cyrille, B. Dieny, B. Ocker, J. Langer, and W. Maas, Spin

torque driven excitations in a synthetic antiferromagnet, *Appl. Phys. Lett.* **96**, 072511 (2010).

[16] G. Y. Shi, C. H. Wan, Y. S. Chang, F. Li, X. J. Zhou, P. X. Zhang, J. W. Cai, X. F. Han, F. Pan, and C. Song, Spin-orbit torque in MgO/CoFeB/Ta/CoFeB/MgO symmetric structure with interlayer antiferromagnetic coupling, *Phys. Rev. B* **95**, 104435 (2017).

[17] C. Bi, H. Almasi, K. Price, T. Newhouse-Illige, M. Xu, S. R. Allen, X. Fan, and W. Wang, Anomalous spin-orbit torque switching in synthetic antiferromagnets, *Phys. Rev. B* **95**, 104434 (2017).

[18] R. A. Duine, K.-J. Lee, S. S. P. Parkin, and M. D. Stiles, Synthetic antiferromagnetic spintronics, *Nat. Phys.* **14**, 217 (2018).

[19] R. Cheng, D. Xiao, and A. Brataas, Terahertz Antiferromagnetic Spin Hall Nano-Oscillator, *Phys. Rev. Lett.* **116**, 207603 (2016).

[20] E. V. Gomonay and V. M. Loktev, Spintronics of antiferromagnetic systems (Review Article), *Low Temperature Physics* **40**, 17 (2014).

[21] Ø. Johansen and J. Linder, Current driven spin–orbit torque oscillator: ferromagnetic and antiferromagnetic coupling, *Sci. Rep.* **6**, 33845 (2016).

[22] E. Monteblanco, F. Garcia-Sanchez, D. Gusakova, L. D. Buda-Prejbeanu, and U. Ebels, Spin transfer torque nano-oscillators based on synthetic ferrimagnets: Influence of the exchange bias field and interlayer exchange coupling, *J. Appl. Phys.* **121**, 013903 (2017).

[23] H. Suhl, *Relaxation Processes in Micromagnetics* (Oxford University Press, New York, 2007).

[24] M. Sparks, *Ferromagnetic-relaxation Theory* (McGraw-Hill Press, New York, 1964).

[25] H. Kurebayashi, T. D. Skinner, K. Khazen, K. Olejník, D. Fang, C. Ciccarelli, R. P. Campion, B. L. Gallagher, L. Fleet, A. Hirohata, and A. J. Ferguson, Uniaxial anisotropy of two-magnon scattering in an ultrathin epitaxial Fe layer on GaAs, *Appl. Phys. Lett.* **102**, 062415 (2013).

[26] M. Körner, K. Lenz, R. A. Gallardo, M. Fritzsche, A. Mücklich, S. Facsko, J. Lindner, P. Landeros, and J. Fassbender, Two-magnon scattering in permalloy thin films due to rippled substrates, *Phys. Rev. B* **88**, 054405 (2013).

[27] I. Barsukov, P. Landeros, R. Meckenstock, J. Lindner, D. Spoddig, Z.-A. Li, B. Krumme, H. Wende, D. L. Mills, and M. Farle, Tuning magnetic relaxation by oblique deposition, *Phys. Rev. B* **85**, 014420 (2012).

[28] K. Zakeri, J. Lindner, I. Barsukov, R. Meckenstock, M. Farle, U. von Hörsten, H. Wende, W. Keune, J. Rocker, S. Kalarickal, K. Lenz, W. Kuch, K. Baberschke, and Z. Frait, Spin dynamics in ferromagnets: Gilbert damping and two-magnon scattering, *Phys. Rev. B* **76**, 104416 (2007).

[29] F. Heimbach, T. Stückler, H. Yu, and W. Zhao, Simulation of high k-vector spin wave excitation with periodic ferromagnetic strips, *J. Magn. Magn. Mater.* **450**, 29 (2018).

[30] A. V. Chumak, V. I. Vasyuchka, A. A. Serga, and B. Hillebrands, Magnon spintronics, *Nat. Phys.* **11**, 453 (2015).

[31] J. Chen, F. Heimbach, T. Liu, H. Yu, C. Liu, H. Chang, T. Stückler, J. Hu, L. Zeng, Y. Zhang, Z. Liao, D. Yu, W. Zhao, and M. Wu, Spin wave propagation in perpendicularly magnetized nm-thick yttrium iron garnet films, *J. Magn. Magn. Mater.* **450**, 3 (2018).

[32] S. M. Rezende, R. L. Rodríguez-Suárez, and A. Azevedo, Magnetic relaxation due to spin pumping in thick ferromagnetic films in contact with normal metals, *Phys. Rev. B* **88**, 014404 (2013).

[33] J. Hellsvik, B. Skubic, L. Nordström, and O. Eriksson, Simulation of a spin-wave instability from atomistic spin dynamics, *Phys. Rev. B* **79**, 184426 (2009).

[34] K. Baberschke, Why are spin wave excitations all important in nanoscale magnetism?, *physica status solidi (b)* **245**, 174 (2008).

[35] S. Qiao, S. Yan, S. Kang, Q. Li, Y. Qin, Y. Zhao, Z. Zhang, and S. Li, Magnetization precession by short-wavelength magnon excitations and spin-transfer torque, *Phys. Rev. B* **97**, 024424 (2018).

[36] M. D. Stiles and J. Miltat, in *Spin Dynamics in Confined Magnetic Structures III*, edited by B. Hillebrands, and A. Thiaville (Springer Press, Berlin, Heidelberg, 2006).

[37] D. C. Ralph and M. D. Stiles, Spin transfer torques, *J. Magn. Magn. Mater.* **320**, 1190 (2008).

[38] Z. Zeng, G. Finocchio, and H. Jiang, Spin transfer nano-oscillators, *Nanoscale* **5**, 2219 (2013).

[39] J. M. D. Coey, *Magnetism and Magnetic Materials* (Cambridge University Press, 2009).

[40] E. Monteblanco, D. Gusakova, J. F. Sierra, L. D. Buda-Prejbeanu, and U. Ebels, Redshift and Blueshift Regimes in Spin-Transfer-Torque Nano-Oscillator Based on Synthetic Antiferromagnetic Layer, *IEEE Magnetics Letters* **4**, 3500204 (2013).

[41] S. Cornelissen, L. Bianchini, T. Devolder, J.-V. Kim, W. Van Roy, L. Lagae, and C. Chappert, Free layer versus synthetic ferrimagnet layer auto-oscillations in nanopillars processed from MgO-based magnetic tunnel junctions, *Phys. Rev. B* **81**, 144408 (2010).